\documentclass[aps,prc,twocolumn,superscriptaddress,showpacs,floatfix,nofootinbib]{revtex4}
\usepackage{amsmath,graphicx}
\newcommand{\mean}[1]{\left\langle #1 \right\rangle}
\begin{document}
\preprint{CERN-PH-TH-2005-???, SPhT-t05/?}

\title{Effects of flow fluctuations and partial thermalization on  $v_4$}

\author{Cl\'ement Gombeaud}
\author{Jean-Yves Ollitrault}
\affiliation{
CNRS, URA2306, IPhT, Institut de physique theorique de Saclay, F-91191
Gif-sur-Yvette, France} 
\date{\today}

\begin{abstract}
The second and fourth Fourier harmonic of the azimuthal distribution of
particles, $v_2$ and $v_4$, have been measured in Au+Au collisions at 
the Relativistic Heavy Ion Collider (RHIC). The harmonic $v_4$
is mainly induced from $v_2$ as a higher-order effect. However,  
the ratio $v_4/(v_2)^2$ is significantly larger than predicted by
hydrodynamics. 
Effects of partial thermalization are estimated on the basis of a
transport calculation, and are shown to increase $v_4/(v_2)^2$ by a
small amount. 
We argue that the large value of $v_4/(v_2)^2$ seen experimentally is
mostly due to elliptic flow fluctations. 
However, the standard model of eccentricity fluctuations is unable to 
explain the large magnitude of $v_4/(v_2)^2$ in central collisions. 
\end{abstract}


\maketitle
\section{Introduction}
The azimuthal distribution of particles emitted in 
ultrarelativistic nucleus-nucleus collisions at RHIC is a sensitive
tool in understanding the bulk properties of the matter produced in
these collisions (see \cite{Voloshin:2008dg} for a recent review). 
It is generally written as a Fourier series
\begin{equation}
\frac{dN}{d\phi}\propto 1+2 v_2\cos 2\phi+2v_4 \cos 4\phi+\cdots
\label{dndphi}
\end{equation}
where $\phi$ is the azimuthal angle with respect to the direction of
flow. 
In this paper, we consider analyses done near the center-of-mass
rapidity, so that odd harmonics vanish by symmetry. 
The large magnitude of elliptic flow, $v_2$, suggests that the
lump of matter formed in a Au-Au collision at RHIC is close to 
local thermal equilibrium and expands as a relativistic fluid. 
Elliptic flow is large at high $p_t$ (up to $0.25$ for baryons), 
which motivated the idea to study the higher-order harmonic 
$v_4$~\cite{Kolb:1999it,Kolb:2003zi}.  
Several analyses of $v_4$ have been
reported~\cite{Adams:2003zg,Masui:2005aa,Abelev:2007qg,Huang:2008vd}. 
Experimental results give $v_4\simeq (v_2)^2$, while the ideal-fluid picture 
generally predicts $v_4=\frac{1}{2}(v_2)^2$~\cite{Borghini:2005kd}. 
This discrepancy has not yet been explained. 
In this paper, we investigate the sensitivity of $v_4$ to two effects:
viscous deviations from the ideal-fluid picture
(Sec.~\ref{sec:partial}), and elliptic flow fluctuations 
(Sec.~\ref{sec:fluctuations}).

\begin{figure}
\includegraphics*[width=0.6\linewidth]{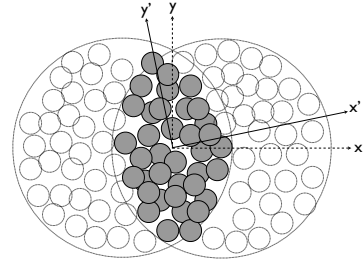}
\caption{Schematic picture of a nucleus-nucleus collision depicted in
  the transverse plane (from~\cite{Alver:2008zz}). The principal axes
  ($x'$ and $y'$) of the area formed by the participants are tilted
  with respect to the reaction plane given by the axes ($x$ and $y$)
 of the transverse plane.}
\label{fig:overlap}
\end{figure}

\section{Ideal hydrodynamics}
We first briefly recall the prediction of relativistic hydrodynamics. 
In this theory, the $\phi$ dependence of particle distribution results
from a similar $\phi$ dependence of the fluid
4-velocity~\cite{Borghini:2005kd,Voloshin:1996nv}:
\begin{equation}
u(\phi)=U\left(1+2 V_2\cos 2\phi+2 V_4\cos 4\phi\cdots\right),
\label{uphi}
\end{equation}
where $\phi$ is the azimuthal angle of the fluid velocity with respect
to the minor axis of the participant ellipse~\cite{Manly:2005zy}
(see Fig.~\ref{fig:overlap}). 
This is due to the fact that the overlap area between the two
colliding nuclei is elliptic, which results in anisotropic pressure
gradients.
For a semi-central Au-Au collision at RHIC, $V_2\sim 4\%$, and one
expects $V_4$ to be of much smaller magnitude, typically $V_4\sim (V_2)^2$. 

The fluid expands, becomes dilute and eventually transforms into
particles. As argued in Ref.~\cite{Borghini:2005kd}, fast particles are
produced where the fluid velocity is maximum, and parallel to the
particle momentum. The resulting momentum distribution is a boosted
thermal distribution. Neglecting quantum statistics (this is justified
in the transverse momentum range where $v_4$ is measured),
the momentum distribution for a given particle of mass $m$ is 
\begin{equation}
\frac{dN}{p_t dp_t d\phi}\propto e^{-p\cdot   u/T}= \exp\left(-\frac{m_t u_0(\phi)-p_t u(\phi)}{T}\right),
\label{dndptdphi}
\end{equation}
where $m_t=\sqrt{p_t^2+m^2}$, $u_0(\phi)=\sqrt{1+u(\phi)^2}$, and
$\phi$ is the azimuthal angle of the particle. 
Inserting Eq.~(\ref{uphi}) into Eq.~(\ref{dndptdphi}), expanding to
leading order in $V_2$, $V_4$ and identifying with Eq.~(\ref{dndphi}),
one obtains~\cite{Borghini:2005kd}  
\begin{eqnarray}
v_2(p_t)&=&\frac{V_2 U}{T}\left(p_t-m_t v\right)\cr
v_4(p_t)&=&\frac{1}{2}v_2(p_t)^2+\frac{V_4 U}{T}\left(p_t-m_t
v\right),
\label{saddlepoint}
\end{eqnarray}
where $v\equiv U/\sqrt{1+U^2}$. 
The higher harmonic $v_4$ is the sum of two contributions: an ``intrinsic''
$v_4$ proportional to the $\cos 4\phi$ term in the fluid velocity 
distribution, $V_4$, and a contribution induced by 
elliptic flow itself, which turns out to be exactly
$\frac{1}{2}(v_2)^2$.   
The latter contribution becomes dominant as $p_t$ increases. 

\begin{figure}
\includegraphics*[width=\linewidth]{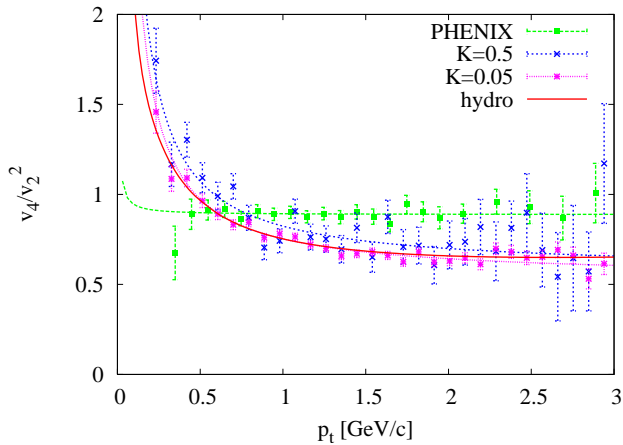}
\caption{(Color online) $v_4/(v_2)^2$ versus $p_t$ in Boltzmann transport theory
  and ideal hydrodynamics for massless particles. 
Solid lines are 2-parameter fits using Eq.~(\ref{fitformula}) over the
  interval $[0.5,2.5]$~GeV/$c$. 
The curves are labeled by the value of the Knudsen number $K$.
Error bars are statistical.
The square dots are results for charged pions from
PHENIX~\cite{Huang:2008vd}, averaged over the centrality interval 20-60\%.}  
\label{fig:fig1}
\end{figure}
In order to confirm these qualitative results, we solve numerically
the equations of ideal relativistic hydrodynamics. 
The fluid is initially at rest. We choose a gaussian initial entropy density
profile, with rms widths $\sigma_x=2$~fm and $\sigma_y=3$~fm.
The equation of state is that of an two-dimensional ideal gas of
massless particles, $s\propto T^2$, for reasons to be explained below. 
The normalization has been fixed in such a way that the
average transverse momentum per particle is $\langle
p_t\rangle=0.42$~GeV/$c$, which is roughly the value for pions in a
central Au-Au collision at RHIC~\cite{Abelev:2008ez}. 
Fig.~\ref{fig:fig1} displays the variation of $v_4/(v_2)^2$ with 
the particle transverse momentum $p_t$. 
For massless particles, $m_t=p_t$ and Eq.~(\ref{saddlepoint}) gives
$v_4/(v_2)^2=0.5+k/p_t$, where $k$ is independent of $p_t$. 
To check the validity of this formula, our numerical results are
fitted over the interval $0.5<p_t<2.5$~GeV/$c$ by the simple formula
\begin{equation}
\frac{v_4(p_t)}{v_2(p_t)^2}=A+B\frac{\langle p_t\rangle}{p_t},
\label{fitformula}
\end{equation}
where we have introduced the average transverse momentum $\mean{p_t}$
in such a way that the coefficient $B$ is dimensionless. 
We refer to $A$ (resp. $B$) as to the induced (resp. intrinsic)
$v_4$. 
We find $A=0.557$ and $B=0.479$. The value of 
$A$ is close to the expected value $0.5$. The small discrepancy is
due to the fact that Eqs.~(\ref{saddlepoint}) are only valid for small
values of $v_2$ and 
$v_4$. This approximation breaks down at the upper end of our fitting
interval, where $v_2(2.5~{\rm GeV/}c)=0.51$. 
%modif in v2
This large value is due to the fact that the equation of state is that
of an ideal gas.%
For large $p_t$, however, 
the intrinsic $V_4$ term in Eq.~(\ref{saddlepoint}) can be neglected,
because it is linear in $p_t$ while the other term is quadratic in
$p_t$. Neglecting this term, the Fourier expansion in
Eq.~(\ref{dndphi}) can be done exactly. This yields  
\begin{equation}
v_{2n}(p_t)=\frac{I_n(x)}{I_0(x)}, 
\label{besselvn}
\end{equation}
where $x=2V_2U(p_t-m_t v)/T$, and $I_n(x)$ is the modified Bessel function. 
Inverting Eq.~(\ref{besselvn}) with $n=1$ and $v_2=0.51$, one 
obtains $x=1.19$. Eq.~(\ref{besselvn}) with $n=2$ then gives 
$v_4/(v_2)^2=0.552$, in better agreement with our numerical result. 

We have systematically investigated the sensitivity of our
hydrodynamical results to initial conditions.
With a smaller initial
eccentricity ($\sigma_x=2$~fm and $\sigma_y=2.5$~fm), the value of
$A$ is closer to $0.5$, as expected from the discussion above. 
We have also repeated the calculation with a more realistic density
profile corresponding to a Au-Au collision at RHIC, obtained using 
an optical Glauber model calculation. We expected
that $B$, which we 
understand as the ``intrinsic'' $v_4$, would be sensitive to the
change in initial conditions, but the changes in both $A$ and $B$ were
insignificant. 

Experimental results are also shown in Fig.~\ref{fig:fig1}. 
The value of $v_4/v_2^2$ is constant, even at relatively low $p_t$:
a fit to these results using Eq.~(\ref{fitformula})
gives $B=0.01\pm 0.04$, compatible with zero.\footnote{Note, however, that STAR
  results for charged 
  particles~\cite{Bai:2007ky}  clearly display an intrinsic $v_4$
  component, although smaller than in our calculation.} 
The other fit parameter is $A=0.89\pm0.02$, significantly 
larger than the value $0.5$ predicted by hydrodynamics. 
Some of the discrepancies between our model calculation and data can
be attributed to the equation of state, which is much softer in QCD
near the transition region than in our hydrodynamical calculation. 
More specifically, the coefficient $B$ representing 
the intrinsic $v_4$ may depend on the equation of state.  
It would be interesting to investigate whether the small value of $B$
seen experimentally can be attributed to the softness of the equation
of state.
On the other hand, our argument leading to $A=\frac{1}{2}$ is
quite general, so that the discrepancy with data cannot be attributed
to the equation of state. 
In this paper, we investigate the possible origins of this
discrepancy. 

\section{Partial thermalization}
\label{sec:partial}

It has been argued~\cite{Bhalerao:2005mm} that if interactions among
the produced particles are not strong enough to produce local thermal
equilibrium, so that the hydrodynamic description breaks down, the
resulting value of $v_4/(v_2)^2$ is higher. This is confirmed by
transport calculations within the AMPT model~\cite{Chen:2004dv}.
This naturally raises the question of how $v_4$ reaches the
hydrodynamic limit~\cite{Lacey:2009xx}. 
We investigate this issue systematically by solving 
numerically a relativistic Boltzmann equation, where the mean free 
path $\lambda$ of the particles can be tuned by varying the elastic
scattering cross section $\sigma$. The degree of thermalization is
characterized by the Knudsen number 
\begin{equation}
K=\frac{\lambda}{R},
\end{equation}
where $R$ is a measure of the system size. 
We consider massless particles moving in the
transverse plane (no longitudinal motion)~\cite{Gombeaud:2007ub}. 
In the limit $K\to 0$, this Boltzmann equation is expected to be
equivalent to ideal hydrodynamics, with the equation of state of a
two-dimensional ideal gas. 
For sake of consistency with our hydrodynamical
calculation, the initial phase space distribution 
of particles is locally thermal: $dN/d^2x d^2
p_t\propto\exp(-p_t/T(x,y))$, where the temperature profile $T(x,y)$ 
is the same as in the hydrodynamical calculation. 
The Knudsen number is normalized as in Ref.~\cite{Gombeaud:2007ub}:
\begin{equation}
K=\frac{4\pi\sqrt{\sigma_x^2+\sigma_y^2}}{N\sigma},
\end{equation}
where $N$ is the total number of particles in the Monte-Carlo
simulation, and $\sigma$ the scattering cross section, which has the
dimension of a length in two dimensions. 
Fig.~\ref{fig:fig1} displays our results for two values of $K$. 
The results for $K=0.05$ are almost identical to the results from 
ideal hydrodynamics, as expected. For $K=0.5$, 
$v_4/(v_2)^2$ is larger, as anticipated in Ref.~\cite{Bhalerao:2005mm}. 
Although the fit formula (\ref{fitformula}) is inspired by
hydrodynamics, the quality of  
the fit is equally good for the Boltzmann calculation.
In particular, the ratio $v_4/(v_2)^2$ quickly 
saturates with increasing $p_t$, which means that the scaling
$v_4\propto (v_2)^2$ still holds if the system does not reach local thermal
equilibrium, as already observed in previous transport
calculations~\cite{Yan:2006bx}.

\begin{figure}
\includegraphics*[width=0.7\linewidth]{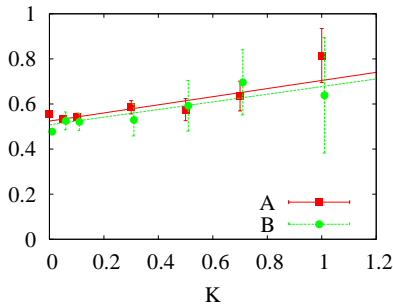}
\caption{(Color online) Variation of the dimensionless fit parameters $A$ and $B$ from
  Eq.~(\ref{fitformula}) with the Knudsen number $K$. Error bars are 
  statistical. Lines are linear fits. 
  The points at $K=0$ are obtained from an independent  
  hydrodynamical calculation and are excluded from the fit. }
\label{fig:fig2}
\end{figure}
The sensitivity of $v_4$ to the Knudsen number $K$ is 
seen more 
clearly in Fig.~\ref{fig:fig2}, which displays the variation of the
fit parameters $A$ and $B$ with $K$. A linear extrapolation of our
Boltzmann results to the limit $K=0$ gives $A=0.524\pm
0.008$ and $B=0.508\pm 0.012$, to be compared with our results from 
ideal hydrodynamics $A=0.557$ and $B=0.479$, in good
agreement\footnote{There is a small residual discrepancy of a few
  percent between Boltzmann and ideal hydrodynamics, which we do not
  understand.}. 

These transport results may be sensitive to the 
choice of initial conditions. We have assumed a locally thermal
momentum distribution. Now, the prediction  $v_4/(v_2)^2$ from
hydrodynamics originates precisely from the assumption that momentum
distributions are thermal in the rest frame of the fluid, see
Eq.~(\ref{dndptdphi}). 
Replacing the exponential in this equation with
a more general function $f(p\cdot u)$ leads to $v_4/(v_2)^2=ff''/(2 f'^2)$. 
With a Levy distribution
$f(x)=(1+x/n/T)^{-n}$, the value of
$v_4/(v_2)^2$ is enhanced by a factor $(1+n)/n$. 
Values of $n$ inferred from $p_t$ spectra of particles produced in p-p
collisions are close to 10~\cite{Adams:2006nd}, which yields a 
slight increase from the prediction of hydrodynamics. 

Realistic values of the Knudsen number $K$, inferred from the
 centrality dependence of $v_2$~\cite{Drescher:2007cd}, are in the
 range $0.3-0.5$ for semi-central collisions. For these values,
Fig.~\ref{fig:fig2} shows that $v_4/(v_2)^2$ is at most $0.6$, still
 significantly below the experimental value $0.9$. 
We conclude that partial thermalization alone cannot
 explain experimental data.

\section{Centrality dependence of $v_4/(v_2)^2$}
\label{sec:data}

\begin{figure}
\includegraphics*[width=\linewidth]{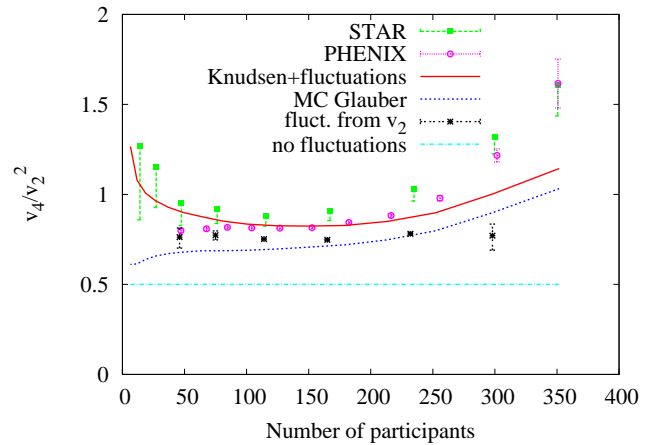}
\caption{(Color online) 
Results from STAR~\cite{YutingPhD} and PHENIX~\cite{lacey} for charged
hadrons produced in Au-Au collisions at 200~GeV per nucleon pair,
versus the number of participant nucleons. 
We have averaged the ratios $v_4/(v_2)^2$ over the intervals 
$1.0<p_t<2.7$~GeV/$c$ for STAR, $1.0<p_t<2.4$~GeV/$c$ for PHENIX.  
Dash-dotted line: prediction from ideal hydrodynamics without flow
fluctuations. 
Stars: with fluctuations inferred from the difference
between $v_2\{2\}$ and $v_2\{{\rm LYZ}\}$, Eq.~(\ref{v4fromv2}).
Dotted line: eccentricity fluctuations from a Monte-Carlo Glauber,
Eq.~(\ref{fluctepsilon}). 
Full line: same, with partial thermalization taken into account,
Eq.~(\ref{fluctv4K}).}
\label{fig:figv4}
\end{figure}

RHIC experiments have  analyzed in detail the centrality dependence
of $v_4/(v_2)^2$. 
Preliminary results from STAR~\cite{YutingPhD} and PHENIX~\cite{lacey}
are presented in Fig.~\ref{fig:figv4}. 
The values of $v_4/(v_2)^2$ are larger than $0.8$ for all 
centralities, and increase up to $1.6$ for central collisions. 
Both experiments observe a similar centrality dependence of
$v_4/(v_2)^2$. STAR obtains values slightly higher than PHENIX. 
This difference may be due to nonflow effects, which are 
smaller for PHENIX than for STAR because the reaction plane detector 
is in a different rapidity window than the central arm 
detector~\cite{Huang:2008vd}. 
Nonflow effects contribute both to $v_2$ and $v_4$. 
We now estimate the order of magnitude of the error on $v_4$. 
We consider for simplicity the case when $v_4$ is analyzed from 
three-particle correlations. 
The corresponding estimate of $v_4$, denoted by 
$v_4\{3\}$~\cite{Borghini:2001vi}, is defined by 
\begin{equation}
v_4\{3\}\equiv \frac{ \mean{\cos(4\phi_1-2\phi_2-2\phi_3)}}{(v_2)^2}
\label{v43}
\end{equation}
where $\phi_j$ are azimuthal angles of outgoing particles and angular
brackets denote an average over triplets of particles belonging to the
same event. In Eq.~(\ref{v43}), $v_2$ must be obtained from another
analysis. 
Nonflow effects arise when particles $1$ and $2$ come from the same
source~\cite{Adams:2003zg}. Assuming that the source flows with the
same $v_2$ as the daughter particles, we obtain
\begin{equation}
\langle \cos(4\phi_1-2\phi_2-2\phi_3)\rangle=v_4(v_2)^2+\delta_{\rm nf}
(v_2)^2,
\end{equation}
where $\delta_{\rm nf}$ is the nonflow correlation. The latter 
can be estimated~\cite{Ollitrault:2009ie} using the 
azimuthal correlation $\delta_{pp}$ measured in proton-proton
collisions~\cite{Adams:2004wz} and scaling it down by the number of
participants:  
$\delta_{\rm nf}=2\delta_{pp}/N_{\rm part}$. 
Dividing by $(v_2)^4$, we obtain the corresponding error on $v_4/(v_2)^2$:
\begin{equation}
\label{systerror}
\delta\left(\frac{v_4}{(v_2)^2}\right)_{\rm
  nf}=\frac{2\delta_{pp}}{N_{\rm part}(v_2)^2}.
\end{equation}
In practice, the analysis is done using the event-plane method rather
than three-particle correlations, but this changes little the
magnitude of nonflow effects~\cite{Ollitrault:2009ie}. 
The error (\ref{systerror}) varies with centrality like $1/\chi^2$,
where $\chi\sim v_2\sqrt{N}$ is the resolution parameter entering the
flow analysis. 
The numerical value $\delta_{pp}=0.0145$ has been used 
in Ref.~\cite{Ollitrault:2009ie} to subtract nonflow effects from $v_2$. It
was obtained by integrating the azimuthal correlation in proton-proton
collisions over $p_t$. 
The error bar on STAR results in Fig.~\ref{fig:figv4} is obtained
using Eq.~(\ref{systerror}) with $\delta_{pp}=0.0145$.  
The agreement with PHENIX is much improved. 
However, this may be a coincidence: in the case of $v_4$, which is 
measured at relatively large $p_t$, nonflow effects are likely to be
larger; on the other hand, nonflow contributions to $v_2$ tend to
increase $v_2$ and decrease the ratio $v_4/(v_2)^2$, which goes in the
opposite direction. 
Finally, we must keep in mind that even with a rapidity gap as in the 
PHENIX analysis, there may be a residual nonflow error of a similar
magnitude.

\section{Flow fluctuations}
\label{sec:fluctuations}

The scaling $v_4=0.5~(v_2)^2$ predicted by ideal hydrodynamics 
only holds for identified particles at a given transverse momentum
$p_t$ and rapidity $y$, for a given initial geometry. 
In order to increase the statistics, however, experimental results for
$v_2$ and $v_4$ are averaged over some of these quantities
{\it before\/} computing the ratio $v_4/(v_2)^2$. 
The averaging process increases the ratio. For instance, the results
shown in Fig.~\ref{fig:fig1} are averaged over a large centrality
interval 20-60\%. 
Even within a narrow centrality class, the initial geometry varies
significantly due to fluctuations in the initial
state~\cite{Aguiar:2001ac,Broniowski:2007ft}
We now discuss the influence of these fluctuations on $v_2$ and $v_4$. 
We assume for simplicity that $v_2$ and $v_4$ are analyzed  using two-particle
correlations and three-particle correlations, respectively. 
The case where the analysis is done using the event-plane method is 
more complex and will be discussed in Sec.~\ref{sec:methods}. 
The estimate of $v_2$ from two-particle correlations is denoted by 
$v_2\{2\}$ and defined by $v_2\{2\}^2\equiv\mean{\cos(2\phi_1-2\phi_2)}$. 
If $v_2$ fluctuates within the sample of events,
$\mean{\cos(2\phi_1-2\phi_2)}=\mean{(v_2)^2}$.  
Similarly, if $v_4$ and $v_2$ fluctuate, 
$\mean{\cos(4\phi_1-2\phi_2-2\phi_3)}=\mean{v_4(v_2)^2}$. 
We thus obtain 
\begin{equation}
\frac{v_4\{3 \}}{v_2\{2 \}^2}=
\frac{\langle v_4 (v_2)^2\rangle}{\langle
(v_2)^2\rangle^2}=\frac{1}{2}
\frac{\langle (v_2)^4\rangle}{\langle (v_2)^2\rangle^2},
\label{fluctv4}
\end{equation}
where, in the last equality, we have assumed that the prediction of
hydrodynamics $v_4=(v_2)^2/2$ holds for a given value of $v_2$. 
If $v_2$ fluctuates, $\langle (v_2)^4\rangle>\langle (v_2)^2\rangle^2$,
which shows that elliptic flow fluctuations increase the observed
$v_4/(v_2)^2$. 
We now estimate quantitatively the magnitude of these
fluctuations. 

\subsection{Flow fluctuations from $v_2$ analyses}
\label{sec:fluctfromv2}

The magnitude of $v_2$ fluctuations can be inferred from the 
difference between estimates of $v_2$, which is dominated
by flow fluctuations except for very peripheral
collisions~\cite{Ollitrault:2009ie}. 
The estimate from 2-particle correlations, $v_2\{2\}$, gives directly
$\mean{(v_2)^2}$, while the estimate of $v_2$ from 4-particle cumulants,
denoted by $v_2\{4\}$, involves $\mean{(v_2)^4}$~\cite{Miller:2003kd}: 
\begin{equation}
v_2\{4\}^4\equiv 2\mean{(v_2)^2}^2-\mean{(v_2)^4}. 
\label{v24}
\end{equation}
Inverting this relation and inserting into Eq.~(\ref{fluctv4}), 
one obtains an estimate of the effect of $v_2$ fluctuations on $v_4$:
\begin{equation}
\frac{v_4\{3 \}}{v_2\{2 \}^2}=\frac{1}{2}
\left(2-\left(\frac{v_2\{4\}}{v_2\{2\}}\right)^4\right).
\label{v4fromv2}
\end{equation}
We use $v\{2\}$ from \cite{Adams:2004bi}; instead of $v_2\{4\}$, we
use the more recent measurement $v_2\{{\rm LYZ}\}$ using Lee-Yang 
zeroes~\cite{Abelev:2008ed,Bhalerao:2003xf}, which is expected to have
a similar sensitivity to flow fluctuations. 
Data on $v_2\{{\rm LYZ}\}$ are only available for semi-central
collisions. The resulting prediction for $v_4/(v_2)^2$ is shown in
Fig.~\ref{fig:figv4}. The agreement with data is much improved when
fluctuations are taken into account. 
We have checked numerically that our results do not change
significantly if nonflow effects are subtracted from $v_2\{2\}$ using
the parametrization introduced in Ref.~\cite{Ollitrault:2009ie}.

\subsection{Flow fluctuations from eccentricity fluctuations}
\label{sec:fluctfromecc}

Since there are
no data on $v_2\{{\rm LYZ}\}$ for the most central and peripheral
bins, we need a model of $v_2$ fluctuations to cover the whole
centrality range. 
We use the standard model of eccentricity 
fluctuations~\cite{Manly:2005zy,Miller:2003kd}. 
The idea is that the overlap area between the colliding nuclei 
(see Fig.~\ref{fig:overlap}) is not smooth: 
positions of nucleons within the nucleus fluctuate from one event to
another, even for a fixed impact parameter. 
Therefore, the participant eccentricity, $\epsilon_{\rm PP}$, which is
the eccentricity of the ellipse defined by the positions of
participant nucleons, also fluctuates. 
Assuming that $v_2$ in a given event scales like $\epsilon_{\rm PP}$, 
Eq.~(\ref{fluctv4}) gives 
\begin{equation}
\frac{v_4\{3 \}}{v_2\{2 \}^2}=\frac{1}{2}
\frac{\langle\epsilon_{\rm PP}^4\rangle}{\langle \epsilon_{\rm PP}^2\rangle^2}.
\label{fluctepsilon}
\end{equation}
We estimate this quantity using the Monte-Carlo Glauber
model~\cite{Miller:2007ri} provided by the PHOBOS
collaboration~\cite{Alver:2008aq}. 
In each event, the participant eccentricity is defined by 
\begin{equation}
\epsilon_{PP}=\frac{\sqrt{(\sigma_y^2 -\sigma_x^2)^2+4\sigma_{xy}^2}}{\sigma_x^2+\sigma_y^2}
\end{equation}
 where $\sigma_x^2=\left\{ x^2 \right\}-\left\{ x \right\}^2$ and
 $\sigma_{xy}=\left\{ xy \right\}-\left\{ x \right\} \left\{ y \right\}$, 
 and $\{\cdots\}$ denotes event-by-event averages over participant
 nucleons.
Each participant is given a 
weight proportional to the number of particles it creates:
\begin{equation}
w=(1-x) + x N_{\rm coll/part},
\end{equation}
where $N_{\rm coll/part}$ is the number of
binary collisions of the nucleon. 
The sum of weights scales like the multiplicity:
\begin{equation}
\frac{dN_{ch}}{d\eta}=n_{pp}\left[
  (1-x)\frac{N_{\rm part}}{2}+xN_{\rm coll}\right ].
\label{mult}
\end{equation}
 where $N_{\rm part}$ and $N_{\rm coll}$ are respectively the number
 of participants and of binary collisions of the considered event. 
 We choose the value $x=0.13$ which best describes the charged hadron
 multiplicity observed experimentally~\cite{Alver:2008aq}. 
 We define the centrality according to the multiplicity
 (\ref{mult}). We evaluate  eccentricity fluctuations in centrality
 classes containing $5\%$ of the total number of events. 
%We have checked that a larger bin size does not significantly change
%our results. 

Our results are presented in Fig.~\ref{fig:figv4}. 
For peripheral and semi-central collisions, the estimates from
eccentricity fluctuations are 
in good agreement with the earlier estimate from the difference 
between $v_2$ analyses, in line with the observation that this
difference is mostly due to eccentricity  
fluctuations~\cite{Ollitrault:2009ie}. 
For the most central bin, however, eccentricity fluctuations only 
increase $v_4/(v_2)^2$ by a factor 2, while a factor 3 would be needed
  to match STAR and PHENIX data. 
This factor 2 can be simply understood. For central collisions, 
eccentricity fluctuations are well described by  a two-dimensional
gaussian distribution~\cite{Voloshin:2007pc}:
\begin{equation}
\frac{dN}{d\epsilon_{\rm PP}}=\frac{\epsilon_{\rm PP}}{\sigma^2}\exp\left(-\frac{\epsilon_{\rm PP}^2}{2\sigma^2}\right). 
\end{equation}
This implies
$\langle\epsilon_{\rm PP}^4\rangle/\langle\epsilon_{\rm PP}^2\rangle^2=2$. 

We now combine the effects of flow fluctuations and 
partial thermalization, discussed in Sec.~\ref{sec:partial}. 
We take partial thermalization into account using the linear fit to
the coefficient $A$ from Fig.~\ref{fig:fig2}: 
\begin{equation}
\frac{v_4}{(v_2)^2}=\frac{1}{2}+0.18~K.
\label{knudsen}
\end{equation}
This modifies Eq.~(\ref{fluctepsilon}) into the following equation:
\begin{equation}
\frac{v_4\{3 \}}{v_2\{2 \}^2}=\left(\frac{1}{2}+0.18~K\right)
\frac{\langle\epsilon_{\rm PP}^4\rangle}{\langle \epsilon_{\rm PP}^2\rangle^2}.
\label{fluctv4K}
\end{equation}
The value of $K$ can be evaluated using the centrality dependence of
elliptic flow. We borrow our estimates from
Ref.~\cite{Drescher:2007cd}. This study has recently been corrected
and refined \cite{Nagle:2009ip}, but the resulting
estimates of $K$ differ little from the original ones. 
Results are shown in Fig.~\ref{fig:figv4}.
Partial thermalization is a small effect. 
Agreement with data is significantly improved for semicentral
collisions, not for central collisions. 
For peripheral collisions, our calculation overshoots PHENIX data. 
Note that Eq.~(\ref{knudsen}) was derived using the results of a
Boltzmann transport calculation, which only applies to a dilute
gas. With a realistic, soft equation of state, the coefficient in
front of $K$ could be different.

\subsection{A toy model of Gaussian flow fluctuations}
\label{sec:gaussianfluct}

\begin{figure}
\includegraphics*[width=\linewidth]{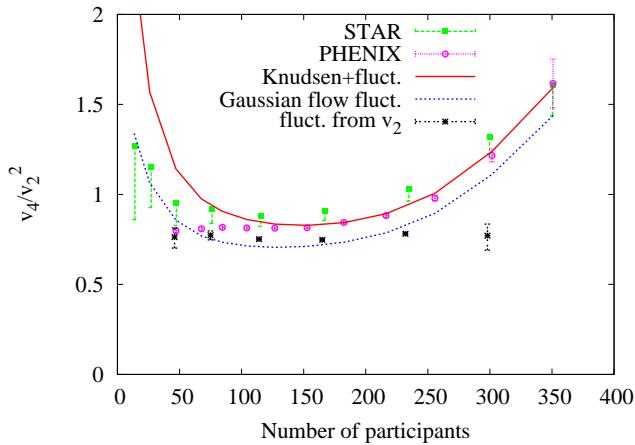}
\caption{(Color online) 
Results using a toy model of gaussian $v_2$ fluctuations. 
STAR and PHENIX data as in Fig.~\ref{fig:figv4}. 
Dashed line: ideal hydrodynamics+gaussian flow fluctuations.
Full line: gaussian flow 
fluctuations and partial thermalization.} 
\label{fig:toy}
\end{figure}

In order to illustrate the sensitivity of $v_4$ to the statistics of
$v_2$ fluctuations, we finally consider a toy model where the
distribution of $v_2$ at fixed impact parameter $b$ is Gaussian:
\begin{equation}
\frac{dN}{dv_2}=\frac{1}{\sigma_v\sqrt{2\pi}}\exp\left(-\frac{(v_2-\kappa\epsilon_{\rm
    RP}(b))^2}{2\sigma_v^2}\right), 
\label{toy}
\end{equation}
where $\epsilon_{\rm RP}$ is the reaction-plane eccentricity obtained
using an optical Glauber model (smooth initial density profile), and
$\kappa$ a proportionality constant. 
We assume that $\sigma_v$ scales like $N_{\rm
  part}^{-1/2}$~\cite{Bhalerao:2006tp}, as 
generally expected for initial state fluctuations, and we adjust the
proportionality constant so as to match the difference between
$v_2\{2\}$ and $v_2\{4\}$ for midcentral collisions. 
The result is displayed in Fig.~\ref{fig:toy}. 
We have checked that a similar result is obtained if we use the
eccentricity from the Color-Glass condensate~\cite{Lappi:2006xc}
instead of the Glauber eccentricity. 
For semicentral and peripheral collisions, this model 
is reasonably close to the standard model of eccentricity
fluctuations. For central collisions, however, results are very 
different, because one-dimensional gaussian fluctuations satisfy 
$\mean{(v_2)^4}/\mean{(v_2)^2}^2=3$ for central collisions, instead of $2$
for eccentricity fluctuations, which are two-dimensional. 
The toy model is in very good agreement with data once 
partial thermalization is taken into account using 
Eq.~(\ref{fluctv4K}). 
However, it lacks theoretical foundations: 
we do not know any microscopic picture that would produce such
gaussian fluctuations. 

\section{Fluctuations and flow methods}
\label{sec:methods}

\begin{figure}
\includegraphics*[width=0.8\linewidth]{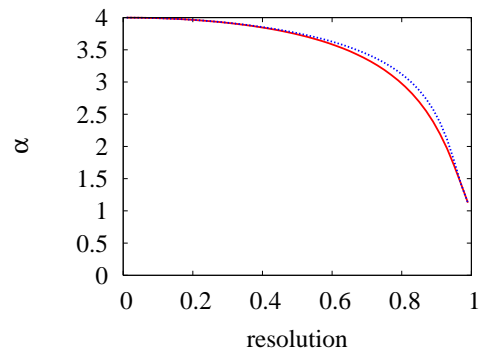}
\caption{Effect of fluctuations on $v_4\{{\rm EP}\}/v_2\{{\rm EP}\}^2$. 
The parameter $\alpha$, defined in Eq.~(\ref{defalpha}), is plotted
versus the resolution of the event-plane for elliptic flow. 
The solid curve is the usual case where the event-plane consists of two
subevents; the dotted curve is the case where the event-plane consists
of only one subevent~\cite{Alver:2008zz}.}
\label{fig:alpha}
\end{figure}

In practice,  $v_2$ and $v_4$ are analyzed
using the event-plane
method~\cite{Poskanzer:1998yz,Ollitrault:1997di}. The 
corresponding estimates are denoted by $v_2\{{\rm EP}\}$ and
$v_4\{{\rm EP}\}$. 
In this Section, we argue that flow fluctuations have almost the same
effect on $v_4\{{\rm EP}\}$ as on $v_4\{3\}$. 
We limit our study to  small fluctuations for simplicity, in the same
spirit as in Ref.~\cite{Ollitrault:2009ie}. 
We write $v_2=\langle v_2\rangle+\delta v$, with $\langle\delta
v\rangle=0$ and $\langle\delta v^2\rangle=\sigma_v^2$, where  
$\sigma_v$ characterizes the magnitude of flow fluctuations.
Expanding
Eq.~(\ref{fluctv4}) to leading order in $\sigma_v$, we obtain
\begin{equation}
\frac{v_4\{3 \}}{v_2\{2 \}^2}=
\frac{1}{2}\left(1+
4\frac{\sigma_v^2}{\langle v_2\rangle^2}\right).
\label{smallfluct}
\end{equation}
Similarly, one can write 
\begin{equation}
\frac{v_4\{{\rm EP}\}}{v_2\{{\rm
    EP}\}^2}=\frac{1}{2} 
\left(1+\alpha\frac{\sigma_v^2}{\langle v_2\rangle^2}\right),
\label{defalpha}
\end{equation}
where $\alpha$ depends on the reaction plane resolution. A similar
parametrization has been introduced for the fluctuations of
$v_2\{{\rm EP}\}$~\cite{Alver:2008zz}. The expression of $\alpha$ is 
derived in Appendix~\ref{sec:appendix} using the same methods as
in Ref.~\cite{Ollitrault:2009ie}. 
Fig.~\ref{fig:alpha} displays the variation of $\alpha$ with the
event-plane resolution for elliptic flow. 
One sees that $\alpha<4$, which means that 
the effect of fluctuations is always smaller for $v_4\{{\rm EP}\}$
than for $v_4\{3\}$; this is confirmed by the experimental
observation $v_4\{3\}>v_4\{{\rm EP}\}$~\cite{Adams:2003zg}. 
The resolution is 1 when the reaction plane is reconstructed
exactly. In this limit, $v_2\{{\rm EP}\}=\langle v_2\rangle$, 
$v_4\{{\rm EP}\}=\frac{1}{2}\langle (v_2)^2\rangle$, which implies $\alpha=1$. In
practice, however, the maximum resolution for mid-central collisions
is  $0.84$ for STAR~\cite{Adams:2004bi} and $0.74$ for
PHENIX~\cite{Huang:2008vd}. In the case of PHENIX, 
$\alpha$ is larger than $3.2$ for all centralities, which means that 
the effect of fluctuations is decreased at most by $20\%$ compared to
our estimates in the previous section.

\section{Discussion}

We have shown that experimental data on $v_4$ are rather well
explained by combining the prediction $v_4=\frac{1}{2}(v_2)^2$ from
ideal hydrodynamics with elliptic flow fluctuations.  
If this scenario is correct, then $v_4/(v_2)^2$ should be independent of
particle species and rapidity for fixed $p_t$ and centrality. 
This is confirmed by preliminary results from PHENIX, which give the
same value for pions, kaons and protons~\cite{Huang:2008vd}. 
Ideal hydrodynamics, which fails to describe $v_2(p_t)$ for
$p_t>1.5$~GeV/c, seems to describe well $v_4/v_2^2$ at least up to
$p_t\sim 3$~GeV/c. 

Note that our scenario does not support the picture of hadron
formation through quark coalescence at large
$p_t$~\cite{Molnar:2003ff}.
We find values of $v_4/v_2^2$ below 1 as a result of the hydrodynamic
expansion, which is believed to take place in the quark phase. 
But coalescence requires that $v_4/(v_2)^2$ for the underlying
quark distribution is much higher, around 2~\cite{Kolb:2004gi}. 
%Indeed, quark coalescence requires that 
%$v_4/(v_2)^2$ for the underlying quark distribution is around 2, 
%significantly larger than the observed $v_4/(v_2)^2$ for 
%hadrons~\cite{Kolb:2004gi}. Now, 
%the model calculation presented in this paper is below  
%the data for hadrons; the discrepancy would be much worse with the 
%underlying quark distribution. 

The centrality dependence of $v_4$ offers a sensitive probe of the
mechanism underlying flow fluctuations. 
Eccentricity fluctuations have been shown to explain quantitatively
$v_2$ data in Au-Au and Cu-Cu collisions. We find that they also
explain most of the results on $v_4$ for peripheral and semi-central
collisions. However, they are unable to explain the steep rise of
$v_4/(v_2)^2$ for the most central bins, which is clearly seen both by
STAR and PHENIX.  
Data suggest that $\mean{(v_2)^4}/\mean{(v_2)^2}^2\simeq 3$ for the most
central bin, while eccentricity fluctuations give 2. 
Impact parameter fluctuations only increase $v_4/v_2^2$ by a few percent.
We cannot exclude a priori that the large experimental value is due to
large errors in the extraction of $v_4$: if we multiply the nonflow
error estimated in 
Sec.~\ref{sec:data} by a factor 4, data agree with our calculation for
central collisions; however, the agreement is spoilt for peripheral
collisions. It therefore seems unlikely that the discrepancy is solely
due to nonflow effects. 
These results suggest that initial state fluctuations do not reduce to  
eccentricity fluctuations, as recently shown by a study of transverse
momentum fluctuations~\cite{Broniowski:2009fm}. 
Interestingly, the direct measurement of $v_2$ fluctuations attempted 
by PHOBOS~\cite{Alver:2007qw}, which agrees with the prediction from 
eccentricity fluctuations, does not extend to the most central bin.

An independent confirmation that $\mean{(v_2)^4}/\mean{(v_2)^2}^2\simeq 3$
for central collisions could be obtained from the 4-particle cumulant
analysis. Interestingly, there is no published value of $v_2\{4\}$ for
the most central bin: the reason is probably that  $v_2\{4\}$ cannot
be defined using Eq.~(\ref{v24}), because the right-hand side is
negative. This  indicates that $\mean{(v_2)^4}/\mean{(v_2)^2}^2>2$. 
It would be interesting to repeat the cumulant analysis for central
collisions, and to scale the right-hand side of  Eq.~(\ref{v4fromv2}) by
$v_2\{2\}^4$. The ratio should be around $-1$ if
$\mean{(v_2)^4}/\mean{(v_2)^2}^2\simeq 3$. 
This would give invaluable information on the mechanism driving elliptic
flow fluctuations.

\appendix
\section{Effect of fluctuations on the event-plane $v_4$}
\label{sec:appendix}
In this Appendix, we derive the expression of $\alpha$ in
Eq.~(\ref{defalpha}). This parameter measures the effect of
fluctuations on $v_4/(v_2)^2$ when flow is analyzed using the
event-plane method. 
The event plane $v_4$ is defined by 
\begin{equation}
v_4\{{\rm EP}\}\equiv \frac{\langle\cos 4(\phi-\Psi_R)\rangle}{R_4},
\label{defv4ep}
\end{equation}
where $\phi$ is the azimuthal angle of the particle, $\Psi_R$ is the
angle of the event plane, and $R_4$ is the event-plane resolution in
the fourth harmonic. 
Using Eq.~(\ref{defv4ep}), the relative variation of $v_4/(v_2)^2$ due to
eccentricity fluctuations can be decomposed as the sum of three contributions
\begin{equation}
\frac{\delta (v_4/(v_2)^2)}{(v_4/(v_2)^2)}=
\frac{\delta \langle\cos 4(\phi-\Psi_R)\rangle}{ \langle \cos
  4(\phi-\Psi_R)\rangle}
-\frac{\delta R_4}{R_4}-2\frac{\delta v_2}{v_2}. 
\label{fluctnumdenom}
\end{equation}
The first term on the right-hand side is the contribution of
fluctuations to the correlation with the event plane, the second term
is the contribution of fluctuations to the resolution, the last term
is the contribution of fluctuations to $v_2\{{\rm EP}\}$. 
The definition of $\alpha$, Eq.~(\ref{defalpha}), can be rewritten as
\begin{equation}
\frac{\delta (v_4/(v_2)^2)}{(v_4/(v_2)^2)}=\frac{\sigma_v^2}{\mean{v_2}^2}\alpha.
\label{defalpha2}
\end{equation}
The three terms in Eq.~(\ref{fluctnumdenom}) give 
additive contributions to $\alpha$, which we evaluate in turn. 

We start with the correlation with the event-plane. 
The event plane $\Psi_R$ is determined from elliptic
flow~\cite{Poskanzer:1998yz}. 
Even flow harmonics $v_{2n}$ are analyzed by correlating particles with
this event plane: 
$\langle\cos
2n(\phi-\Psi_R)\rangle=v_{2n}{\cal R}_{2n}(\chi)$, 
where the resolution ${\cal R}_{2n}$ is given by~\cite{Ollitrault:1997di}
\begin{equation}
{\cal R}_{2n}(\chi)=\frac{\sqrt{\pi}}{2}e^{-\chi^2/2}\chi\left(
  I_{\frac{n-1}{2}}\left(\frac{\chi^2}{2}\right)+I_{\frac{n+1}{2}}\left(\frac{\chi^2}{2}\right) \right),
\label{resolution}
\end{equation}
where $\chi$ is the resolution parameter, which is estimated using the
correlation between two subevents.
For $n=2$, this equation reduces to 
\begin{equation}
{\cal R}_{4}(\chi)=\frac{e^{-\chi^2}-1+\chi^2}{\chi^2}.
\label{resolution4}
\end{equation}
These relations are derived neglecting flow fluctuations. 
If $v_2$ fluctuates, the resolution parameter $\chi$ scales like $v_2$, 
$\chi=rv_2$. Assuming in addition that $v_4$ scales like $(v_2)^2$, the
relative change due to fluctuations is, to leading order in $\sigma_v$, 
\begin{eqnarray}
\frac{\delta \langle\cos 4(\phi-\Psi_R)\rangle}{ \langle \cos
  4(\phi-\Psi_R)\rangle}&=&
\frac{\sigma_v^2}{2}\frac{\frac{d^2}{(dv_2)^2}\left((v_2)^2{\cal
 R}_4(rv_2\right))}{\mean{v_2}^2{\cal R}_4(r\mean{v_2})} \cr
&=&
  \frac{\sigma_v^2}{2\mean{v_2}^2}\frac{\frac{d^2}{d\chi^2}\left(\chi^2{\cal
  R}_4(\chi\right))}{{\cal R}_4(\chi)},
\label{defflnum}
\end{eqnarray}
where the right-hand side is evaluated for $\chi\equiv r\mean{v_2}$,
the average resolution parameter. 
Using Eq.~(\ref{resolution4}), one obtains
\begin{equation}
\frac{1}{{\cal R}_4(\chi)}
\frac{d^2}{d\chi^2}\left(\chi^2{\cal  R}_4(\chi)\right)=
\frac{2\chi^2(e^{\chi^2}+2\chi^2-1)}{1+e^{\chi^2}(\chi^2-1)}.
\end{equation}
Inserting into Eqs.~(\ref{defflnum}) and (\ref{fluctnumdenom}), and
identifying with Eq.~(\ref{defalpha2}), we obtain the  
contribution to $\alpha$ from the correlation with the event plane:
\begin{equation}
\alpha_{\rm
  ep}=\frac{\chi^2(e^{\chi^2}+2\chi^2-1)}{1+e^{\chi^2}(\chi^2-1)}.
\label{alphaep}
\end{equation}

We now evaluate the second term in Eq.~(\ref{fluctnumdenom}), namely,
the shift in the resolution from fluctuations. The resolution is
defined as $R_4\equiv {\cal R}_4(\chi^{\rm exp})$, where $\chi^{\rm exp}$ is
determined from the correlation between subevents. 
Flow fluctuations shift the estimated resolution. Writing 
$\chi^{\rm exp}=\chi+\delta\chi$, one obtains, to leading order in
$\delta\chi$, 
\begin{equation}
\frac{\delta R_4}{R_4}=\frac{\chi {\cal R}_4'(\chi)}{{\cal R}_4(\chi)}
\frac{\delta\chi}{\chi}.
\label{deltar4}
\end{equation}
Eq.~(\ref{resolution4}) gives
\begin{equation}
\frac{\chi {\cal R}_4'(\chi)}{{\cal
    R}_4(\chi)}=\frac{2(e^{\chi^2}-\chi^2-1)}{1+e^{\chi^2}(\chi^2-1)}. 
\label{rprime4}
\end{equation}
The shift in the resolution to fluctuations is given by Eq.~(A7) of 
Ref.~\cite{Ollitrault:2009ie}
\begin{equation}
\frac{\delta\chi}{\chi}
=\frac{\sigma_v^2}{2\mean{v}^2}
\left(1-2\chi_s^2+\frac{4 i_1^2}{i_0^2-i_1^2}\right).
\label{denom1}
\end{equation}
where $i_{0,1}$ is a shorthand notation for $I_{0,1}(\chi_s^2/2)$, and
$\chi_s$ denotes the resolution parameter of a subevent:
$\chi_s=\chi/\sqrt{2}$ in the usual case when the event plane consists
of two subevents~\cite{Poskanzer:1998yz}, and $\chi_s=\chi$ if the
event plane has only one subevent~\cite{Alver:2008zz}. 
Inserting Eqs.~(\ref{rprime4}) and (\ref{denom1}) into (\ref{deltar4})
and (\ref{fluctnumdenom}), and
identifying with Eq.~(\ref{defalpha2}), we obtain the  
contribution to $\alpha$ from the resolution:
\begin{equation}
\alpha_{\rm res}=
\frac{e^{\chi^2}-\chi^2-1}{1+e^{\chi^2}(\chi^2-1)}\left(1-2\chi_s^2+\frac{4
  i_1^2}{i_0^2-i_1^2}\right) 
\label{alphares}
\end{equation}

Finally, the third term in Eq.~(\ref{fluctnumdenom}) is 
\begin{equation}
2\frac{\delta v_2}{v_2}=\frac{\sigma_v^2}{\mean{v_2}^2}(\alpha_{v_2}-1)
\end{equation}
where $\alpha_{v_2}$ is given by Eq.~(23) of Ref.~\cite{Ollitrault:2009ie}:
\begin{equation}
\alpha_{v_2}=2
-\frac{I_0-I_1}{I_0+I_1}\left(2\chi^2-2\chi_s^2+\frac{4
   i_1^2}{i_0^2-i_1^2}\right),
\label{alpha2ep}
\end{equation}
where $I_{0,1}$ is a shorthand notation for $I_{0,1}(\chi^2/2)$. 

The final result is obtained by summing the three contributions from
Eqs.~(\ref{alphaep}), (\ref{alphares}) and (\ref{alpha2ep}):
\begin{equation}
\alpha=\alpha_{\rm ep}-\alpha_{\rm res}-(\alpha_{v_2}-1). 
\end{equation}
The limit of low resolution $\chi\to 0$ (resp. high resolution
$\chi\to\infty$) is $\alpha_{\rm ep}=6$ (resp. 1), $\alpha_{\rm res}=1$
(resp. 0), $\alpha_{v_2}=2$ (resp. 1), $\alpha=4$ (resp. 1). 

\section*{Acknowledgments}

We thank Y. Bai, S. Huang and R. Lacey for sending us
preliminary data from STAR and PHENIX, and C. Loizides for
permission to use a figure of Ref.~\cite{Alver:2008zz}. 
We thank F. Gelis, T. Lappi, M. Luzum, J.L. Nagle, H.
Pereira da Costa, R. Snellings and A. Tang for useful
discussions. We are grateful to J.-P. Blaizot and A.M. Poskanzer
for useful comments on the manuscript. 
This work is funded by ``Agence Nationale de la Recherche'' under grant
ANR-08-BLAN-0093-01.

\end{document}